# Tunable Terahertz Amplification Based on Photoexcited Active Graphene Hyperbolic Metamaterials

TIANJING GUO,[1] LIANG ZHU,[2] PAI-YEN CHEN,[2] AND CHRISTOS ARGYROPOULOS[1,*]

[1]*Department of Electrical and Computer Engineering, University of Nebraska-Lincoln, Lincoln, NE, 68588, USA*
[2]*Department of Electrical and Computer Engineering, University of Illinois at Chicago, Chicago, IL 60607, USA*
*\*christos.argyropoulos@unl.edu*

**Abstract:** The efficient amplification and lasing of electromagnetic radiation at terahertz (THz) frequencies is a non-trivial task achieved mainly by quantum cascade laser configurations with limited tunability and narrowband functionality. There is a strong need of compact and efficient THz electromagnetic sources with reconfigurable operation in a broad frequency range. Photoexcited graphene can act as the gain medium to produce coherent radiation at low THz frequencies but its response is very weak due to its ultrathin thickness. In this work, we demonstrate an alternative design to achieve efficient tunable and compact THz amplifiers and lasers with broadband operation based on active THz hyperbolic metamaterials (HMM) designed by multiple stacked photoexcited graphene layers separated by thin dielectric sheets. The hyperbolic THz response of the proposed ultrathin active HMM is analytically and numerically studied and characterized. When the graphene-based HMM structure is periodically patterned, a broadband slow-wave propagation regime is identified, thanks to the hyperbolic dispersion. In this scenario, reconfigurable amplification of THz waves in a broad frequency range is obtained, which can be made tunable by varying the quasi-Fermi level of graphene. We demonstrate that the THz response of the presented tunable THz amplifiers or lasers is controlled by the incident optical pumping (photodoping) and the loaded dielectric materials in the HMM waveguide array, an interesting property that can have great potential for THz amplification, emission, and sensing applications.

## 1. Introduction

Hyperbolic metamaterials (HMM) [1] are anisotropic and uniaxial artificial engineered electromagnetic materials. They have recently attracted intense scientific interest and became popular for engineering and controlling the propagation, scattering, and radiation of light. To date, HMMs have been exploited in a variety of applications, such as negative refraction [2–5], superlensing [1], biosensing [6], super absorbers [7], energy harvesting [7–10], and enhancement of spontaneous emission [11–13] or Purcell effect [14]. Several additional applications of HMM substrates have been discussed in [15]. The two common categories of HMM are composed of either metal-dielectric multilayers or metallic nanorod arrays [1], which can be readily fabricated by chemical vapor deposition (CVD) or electrochemical techniques. Metals play an important role in these configurations, as they provide the desired plasmonic behavior which leads to negative permittivity at optical frequencies, crucial for achieving the hyperbolic dispersion.

Graphene is a typical two-dimensional (2D) nanomaterial [16] made of a single layer of carbon atoms arranged into a hexagonal lattice. It is expected to replace metals at THz and infrared (IR) frequencies, as the crucial building element of photonic devices, due to its relatively low loss plasmonic response and tunable performance at this frequency range. In addition, the CVD-grown graphene monolayer can be an effective building block of layered structures, such as HMMs and metasurfaces [17–23]. More interestingly, the complex conductivity of graphene, and, as a result, the HMM dispersion, can be tuned as a function of its chemical potential (Fermi energy) over a wide spectrum range [17,24–29], making graphene-based HMM or other metasurfaces to exhibit interesting reconfigurable properties at THz and IR frequencies. Owning to its low density of states, graphene's Fermi energy can be readily tuned via electrostatic gating or chemical doping. In the case of photoexcited graphene, since the relaxation time of intraband transitions is much shorter than the recombination time of electron-hole pairs [30], population inversion can be achieved [31,32]. For a sufficiently strong photoexcitation, THz amplification or gain can be obtained due to the cascaded optical-phonon emission and the interband transition around the Dirac point. In this nonequilibrium scenario, the real part of the dynamic conductivity of graphene becomes negative [31–38].

However, graphene monolayers are one-atom thick and cannot strongly interact with the incident excitation wave, leading to poor gain efficiency. In addition, tunable and high power coherent electromagnetic radiation sources at low THz frequencies are very limited [39]. Typically, quantum cascade lasers are used as coherent sources in this frequency range but they suffer from poor tunability, leading to narrowband operation, and a complicated bulky design [40]. In this paper, we theoretically propose an alternative tunable and compact THz laser, operating in a broad frequency range, based on periodically-patterned graphene-based HMM that are photoexcited by IR/visible pump light. The proposed graphene nanostructure supports slow-light propagation and a nearly-zero energy velocity at THz frequencies, combined with broad frequency tunability. The THz gain of the photoexcited graphene can be boosted in the slow-wave regime, due to the light trapping effect in the proposed patterned graphene-based HMM configuration [9]. The proposed device has an extremely subwavelength thickness, making it an ideal candidate to be used as a compact THz source in the envisioned THz on-chip integrated photonic circuits.

The paper is structured as following. In section 2, we discuss the negative conductivity of graphene at the THz regime (THz gain) and its tunable properties. By using the effective medium approximation (EMA) [41,42], we demonstrate that the hyperbolic dispersion is obtained for the optically-pumped graphene-based HMM. In section 3, we analytically study the enhanced THz lasing effect of the proposed active THz HMM, patterned into waveguide-array structures. In section 4, we present full-wave simulation results to validate the analytical model and to optimize the THz lasing effect via a detailed parametric study. In section 5, we propose a potential application of the presented active HMM in high-resolution and high quality-factor ($Q$-factor) THz sensing. Finally, we draw conclusions in section 6.

## 2. Hyperbolic metamaterials with active and passive graphene layers

When a graphene monolayer is optically pumped, the electron-hole pairs split the Fermi level into two quasi-Fermi levels $E_{Fn} = \varepsilon_F$ and $E_{Fp} = -\varepsilon_F$, respectively, as shown in Fig. 1(a). In this case, the dynamic conductivity of graphene can be modeled by using the nonequilibrium Green's functions described by [43]:

$$\sigma = j\frac{q^2}{\pi\hbar^2}\frac{1}{\omega - j\tau^{-1}}\left[\int_0^\infty \varepsilon\left(\frac{\partial f_1(\varepsilon)}{\partial \varepsilon} - \frac{\partial f_2(-\varepsilon)}{\partial \varepsilon}\right)d\varepsilon\right] - j\frac{q^2}{\pi\hbar^2}(\omega - j\tau^{-1})\int_0^\infty \frac{f_2(-\varepsilon) - f_1(\varepsilon)}{(\omega - j\tau^{-1})^2 - 4\varepsilon^2/\hbar^2}d\varepsilon, \quad (1)$$

where $f_1(\varepsilon) = \left[1 + e^{(\varepsilon - E_{Fn})/k_B T}\right]^{-1}$, $f_2(\varepsilon) = \left[1 + e^{(\varepsilon - E_{Fp})/k_B T}\right]^{-1}$, $q$ is the electron charge, $\varepsilon$ is the energy, $\hbar$ is the reduced Planck's constant, $k_B$ is the Boltzmann constant, $T$ is the temperature, $\omega$ is the angular frequency, and $\tau$ is the relaxation time of charge carriers. The first term of the formula represents the contribution of the intraband transition and the second part represents the contribution of the interband transition. Without optical excitation, a graphene layer can be modeled by the local surface conductivity formula [20,44]:

$$\sigma = \frac{-jq^2 k_B T}{\pi\hbar^2(\omega - j\tau^{-1})}\left(\frac{E_F}{k_B T} + 2\ln(e^{-E_F/(k_B T)} + 1)\right) - \frac{jq^2(\omega - j\tau^{-1})}{\pi\hbar^2}\int_0^\infty \frac{f_D(-\varepsilon) - f_D(\varepsilon)}{(\omega - j\tau^{-1})^2 - 4\varepsilon^2/\hbar^2}d\varepsilon, \quad (2)$$

where $f_D(\varepsilon) = [e^{(\varepsilon - E_F)/(k_B T)} + 1]^{-1}$ is the Fermi-Dirac distribution, and $E_F$ is the Fermi energy or doping level in the passive graphene case. Figure 1(b) demonstrates the calculated frequency dispersion of the real-part conductivity $\text{Re}[\sigma]$ for a graphene monolayer with and without optical pumping (i.e., photodoping) at a temperature of 77 K. Negative real conductivity over a broad frequency range is obtained when the graphene monolayer is pumped with quasi-Fermi level equal to 50 meV, as can be seen in Fig. 1(b) (solid black line). On the other hand, the unpumped (passive) graphene monolayer exhibits positive real conductivity at the same temperature and frequency range [Fig. 1(b), solid blue line]. Reducing the temperature can increase the absolute value of the negative real conductivity in the pumped case, and vice versa. The negative conductivity and the effect of THz gain disappear at high temperatures. The reflectance and transmittance of both pumped and unpumped cases of monolayer graphene are plotted in Fig. 1(c). They are found to have almost identical values with unitary transmission and low reflection along a broad frequency range. This is because a single graphene monolayer does not interact strongly with the incident THz radiation, even if it is photoexcited, mainly due to its ultrathin thickness.

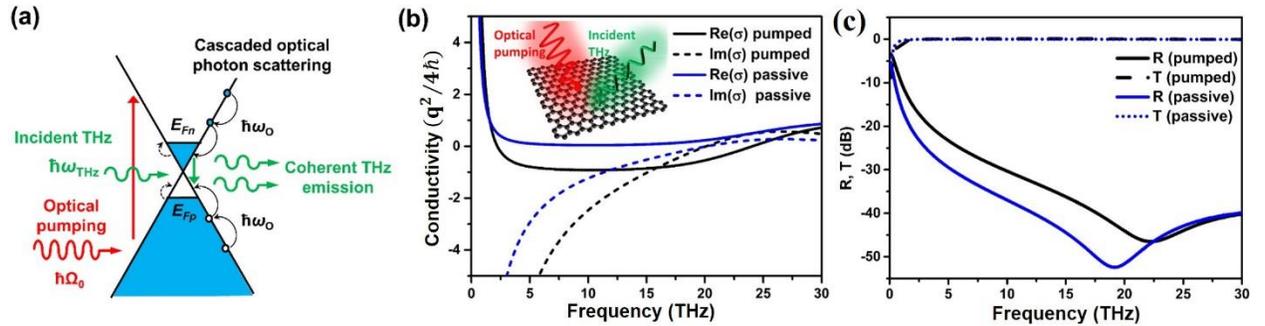

Fig. 1. (a) Schematic of the terahertz gain coherent response by an optically pumped graphene monolayer. (b) Frequency dispersion of the real and imaginary parts of the graphene conductivity with (pumped) and without (passive) optical photoexcitation. Inset: Schematic of the optically pumped graphene process. (c) Computed reflectance and transmittance in dB versus frequency for a graphene monolayer with (pumped) and without (passive) optical photoexcitation under normal incident TM polarized wave excitation.

To increase the light-matter interaction, we investigate a multilayer HMM composed of clustered graphene monolayers and dielectric sheets, schematically depicted in the inset of Fig. 2(a). By applying the EMA theory, this structure can be modeled as an anisotropic uniaxial medium, whose effective relative permittivity $\varepsilon_{\textit{eff}}$ can be expressed in a tensor form: $\varepsilon_{\textit{eff}} = \varepsilon_t(\hat{x}\hat{x} + \hat{y}\hat{y}) + \varepsilon_z \hat{z}\hat{z}$ [1]. The transverse relative permittivity $\varepsilon_t$ is given by: $\varepsilon_t = \varepsilon_t' - j\varepsilon_t'' = \varepsilon_d - j\dfrac{\sigma}{\omega \varepsilon_0 d}$ [20,41], where $\varepsilon_d$ and $d$ are the permittivity and thickness of each dielectric layer, and $\sigma$ is the conductivity of graphene given by either Eq. (1) or (2). The longitudinal relative permittivity $\varepsilon_z$ is equal to $\varepsilon_d$ since graphene has negligible thickness compared to the dielectric layers. For both active and passive graphene monolayers, the hyperbolic dispersion is obtained at certain frequencies, where the real part of the transverse effective permittivity $\varepsilon_t'$ is negative, while $\varepsilon_z$ is always positive. Note that only transverse magnetic (TM) polarized waves can exhibit hyperbolic dispersion, as reported in [20].

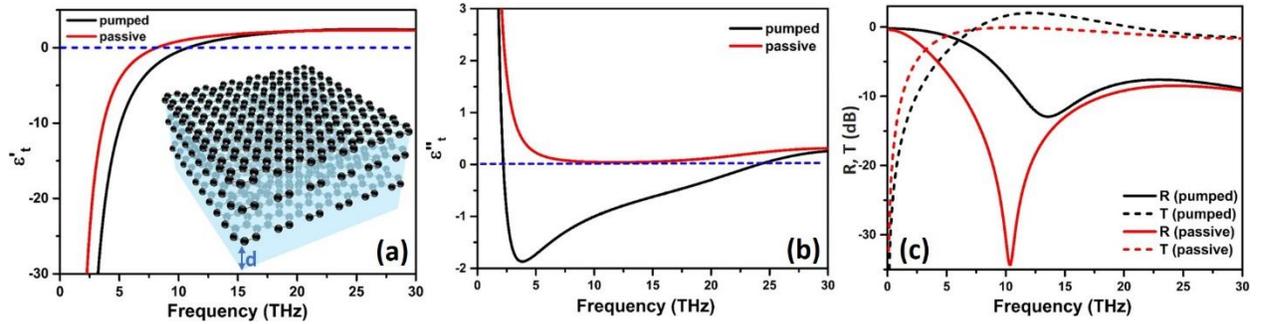

Fig. 2. (a) Real and (b) imaginary parts of the effective transverse relative permittivity $\varepsilon_t = \varepsilon_t' - j\varepsilon_t''$ of the pumped and unpumped graphene HMM multilayer structure with a schematic shown in the inset. The graphene layers are seperated by dielectric slabs with equal thicknesses $d$ and permittivities $\varepsilon_d$. (c) Computed reflectance and transmittance in dB versus frequency of the HMM based on active (pumped) and passive graphene layers under normal incident TM polarized wave excitation.

Figures 2(a,b) represent the real and imaginary parts of the effective transverse relative permittivity, respectively, for the THz HMM made of graphene, shown in the inset of Fig. 2(a), with $\varepsilon_d = 2.2$ and $d = 0.1$ μm. Since graphene is transparent to the near-IR and visible pump waves, without loss of generality, we assume that each graphene monolayer is uniformly photodoped to have the same quasi-Fermi level. Figure 2(b) shows that the photopumped graphene-based HMM has a negative imaginary part of the effective transverse relative permittivity $\varepsilon_t''$ (gain) over a broad frequency range, where $\varepsilon_t'$ and $\varepsilon_z$ have opposite signs (i.e., hyperbolic dispersion). Figure 2(c) presents the simulated transmittance (dashed lines) and reflectance (solid lines) for the graphene-based HMM, composed of 10 graphene sheets separated by 0.1 μm-thick silica ($\varepsilon_d = 2.2$). Low transmittance is obtained in the hyperbolic-dispersion region for both passive (e.g., for frequencies < 8.1 THz) and active (for frequencies < 11.2 THz) graphene HMMs. On the contrary, high transmittance is obtained in the elliptical-dispersion region for both cases. It is evident from Fig. 2(c) that the active graphene HMM can exhibit a transmittance greater than one or, effectively, negative

absorptance, also known as gain. Nonetheless, the THz gain response of this device is still rather weak, although stronger compared to the interaction in the case of monolayer graphene, as was shown in Fig. 1(c). To significantly enhance the THz lasing effect of graphene, in the following we investigate a patterned graphene-based HMM configuration.

## 3. THz amplification using photoexcited patterned graphene-based hyperbolic metamaterials

In this section, we study a patterned graphene-based HMM with schematic shown in Fig. 3(a), which supports slow-wave modes in the THz region. Slowing light may promote stronger light–matter interaction and, as a result, boost the amplification of THz waves. The proposed slow-wave HMM structure is built by a periodic array of air-HMM-air waveguides. The dispersion relationship between angular frequency $\omega$ and the propagation constant $\beta$ is derived by using the transverse resonance method [10], [45] with suitable periodic boundary conditions. After some mathematical manipulations, it is given by the formula:

$$\tan[k'\frac{P-W}{2}] - \varepsilon_z \frac{k''}{k'} \tanh[k''\frac{W}{2}] = 0, \quad (3)$$

where $k' = \sqrt{k_0^2 \varepsilon_z - (\varepsilon_z/\varepsilon_t)\beta^2}$, $k'' = \sqrt{\beta^2 - k_0^2}$, and $k_0$ is the free-space wave number. Note that $\varepsilon_z$ and $\varepsilon_t$ are respectively the effective longitudinal and transverse permittivity of the proposed multilayered HMM structure presented in the previous section, after being modeled as a homogeneous uniaxial anisotropic medium by using the EMA theory. The height of the proposed multilayered structure is not considered in the dispersion relationship given by Eq. (3). Only the effective permittivity and volume fraction is important in the derivation of this dispersion equation, since the height of the proposed structure is extremely subwavelength. The dispersion diagram of the proposed structure is plotted in Fig. 3(b), showing that the slow-light effect can be achieved, with a near-zero group velocity (i.e., $\upsilon_g = \partial\omega/\partial\beta \approx 0$) at the design frequency ($f \approx 7.5$ THz). Note that the slow-light regime can only be achieved with the structure shown in Fig. 3(a) and this response can only be proven by analytically computing its dispersion equation. In this scenario, the incident electromagnetic radiation can be efficiently trapped and amplified by the proposed THz active plasmonic structure, resulting in a strong THz lasing effect. It is worth mentioning that the frequency of the near-zero group velocity can be tuned by several factors, including: (i) the quasi-Fermi level of graphene, (ii) the geometry of the HMM (e.g. periodicity, trench width, and insulator thickness), and (iii) the dielectric material that constitutes the HMM. In addition, it is interesting that the thickness of the proposed THz lasing device, which is composed of only 10 patterned graphene sheets stacked between dielectric layers (Fig. 3(a)), is deeply-subwavelength $d_1 = 10 \times t = 1 \mu m \approx \lambda/30$, where $t = 0.1$ μm. This is another major advantage of the proposed device towards its integration in THz on-chip circuits.

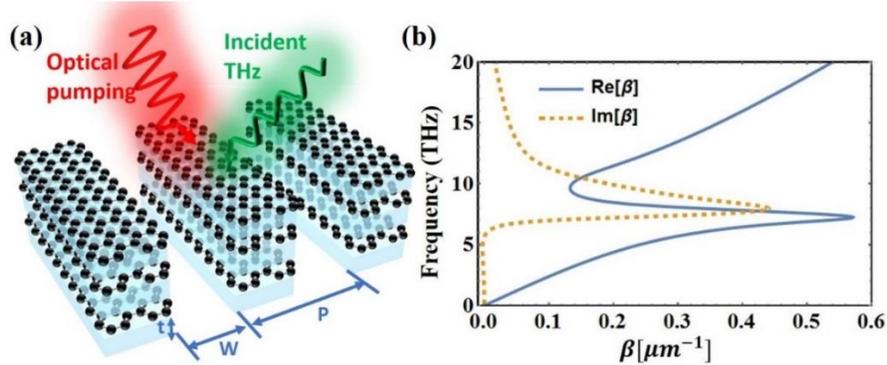

Fig. 3. (a) The proposed patterned HMM structure composed of $N$ = 10 patterned graphene sheets stacked between dielectric layers. $P$ is the period of the device, $W$ is the width of the air slots, and $t$ is the thickness of each dielectric spacer layer. (b) Dispersion diagram of the proposed structure. In this case, the dimension parameters are $P$ =1.9 μm, $W$ = 0.5 μm, $t$ = 0.1 μm and the quasi-Fermi level of the pumped graphene is 50 meV.

Figures 4(a,b) demonstrates the analytical results, computed by using the transmission line model [46] shown in Fig. 5(a), of the reflectance and transmittance, respectively, versus the frequency of the proposed patterned graphene HMM and different quasi-Fermi levels. We find that large forward and backward scattering can be obtained at the lasing point, which coincides with the zero-velocity point. For instance, if the quasi-Fermi level of each graphene is pumped to 50 meV, the lasing peak is obtained at 7.5 THz, agreeing very well with the results presented in Fig. 3(b). It is interesting to note that the lasing peak can be tuned over a wide range of frequencies by varying the quasi-Fermi level of graphene, which can be readily achieved by controlling the intensity of IR or visible pump light. The maximum THz output power is obtained at 5.34 THz, as seen in Fig. 4. This can be explained by using the transmission line model [46] shown in Fig. 5(a). The input impedance $Z_{in}$ at the air/HMM interface is given by:

$$Z_{in} = Z_1 \frac{Z_0 + iZ_1 \tan(\beta d_1)}{Z_1 + iZ_0 \tan(\beta d_1)}, \quad (4)$$

where $Z_1 = (\beta/\omega\varepsilon_0) \times (P-W)$ and $d_1 = 10 \times t = 1\,\mu m$ are the characteristic impedance and total thickness of HMM, respectively, and the impedance of the surrounding free space is $Z_0 = \sqrt{\mu_0/\varepsilon_0}\,P$. The reflectance $R$ and transmittance $T$ are computed by:

$$R = \frac{Z_{in} - Z_0}{Z_{in} + Z_0};$$
$$T = \frac{2Z_{in}}{Z_{in} + Z_0}. \quad (5)$$

Note that the subwavelength height $d_1$ of the proposed structure is included in the transmission/reflection calculations. It can be derived from Eq. (5) that if $Z_{in} = -Z_0$, the transmittance and reflectance of the structure become infinity, provided that the effects of nonlinearity and saturated gain in graphene are negligible. This is clearly depicted in Fig. 5(b), where at the frequency point of 5.34 THz and for $\varepsilon_F = 27$ meV, the optimum lasing condition can be achieved, i.e., $\mathrm{Re}[Z_{in}] = -Z_0$ and $\mathrm{Im}[Z_{in}] = 0$.

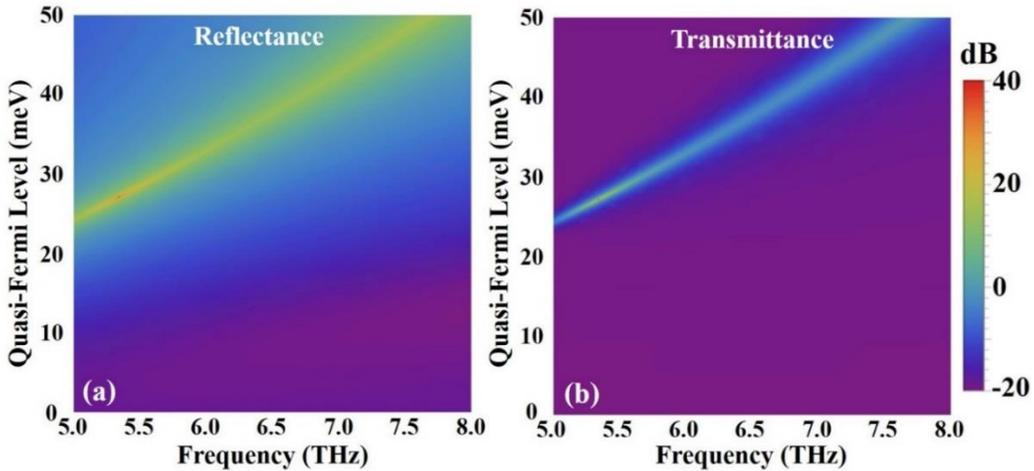

Fig. 4. (a) Reflectance and (b) transmittance in dB computed as a function of the incident frequency and quasi-Fermi level of the THz slow-wave HMM graphene structure shown in Fig. 3(a) with geometrical parameters $P = 1.9\,\mu m$, $W = 0.5\,\mu m$, and $t = 0.1\,\mu m$.

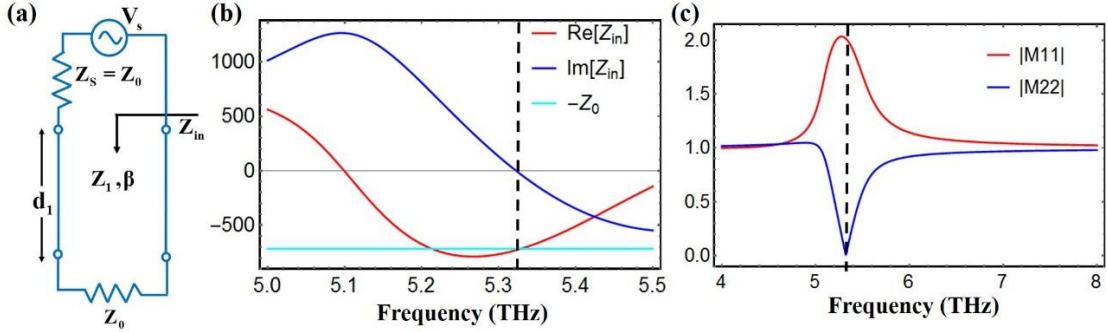

Fig. 5. (a) Transmission line model used to analyze the structure shown in Fig. 3(a). The source impedance is $Z_S = Z_0$, where $Z_0$ is the free space impedance, $Z_{in}$ is the input impedance, $Z_1$ is the characteristic impedance of the proposed HMM structure, $\beta$ is the effective propagation constant, and $d_1$ is the total thickness of the proposed device. (b) Real and imaginary parts of the input impedance computed by the transmission line model. (c) Computed absolute values of the transfer matrix elements $M_{11}$ and $M_{22}$. In all these cases, the quasi-Fermi level is fixed to 27 meV leading to the lasing point shown in Fig. 4.

We also use the transfer matrix method to analyze the presented lasing phenomenon. The transfer matrix that connects the fields between the output and input waves is defined as $|E_i\rangle = M|E_o\rangle$, where the subscripts 'o' and 'i' refer to the electric fields at the output and input surfaces, respectively. The relationship between the elements of the transfer and scattering matrices is described by [47]: $M_{11} = S_{21} - S_{22}S_{11}/S_{12}$, $M_{12} = S_{22}/S_{12}$, $M_{21} = -S_{11}/S_{12}$, $M_{22} = 1/S_{12}$. If the ports 1 and 2 are defined as the excitation and output ports, respectively, $S_{21}$ and $S_{11}$ represent the transmission and reflection coefficients of the proposed structure. Note that $S_{21} = S_{12}$ and $S_{11} = S_{22}$ since the current configuration is symmetric and reciprocal. Figure 5(c) clearly demonstrates that $M_{22} = 0$ at 5.34 THz, which means that the transmission is very high at this frequency point.

## 4. Simulation results

To verify our analytical results, we calculate the gain response of the proposed structure by using COMSOL Multiphysics [48], a finite element method (FEM)-based commercial software that solves Maxwell's equations and computes the reflection and transmission coefficients. Graphene is modelled as a surface current during all our simulations, instead of a bulk permittivity, because this a more accurate and less computationally intensive way to simulate 2D ultrathin materials. The gain coefficient is calculated from the reflectance and transmittance by using the formula: $Gain = 1 - |S_{11}|^2 - |S_{21}|^2$. Note that this formula can be used to compute either gain ($|S_{11}|^2 + |S_{21}|^2 > 1$) or absorptance ($|S_{11}|^2 + |S_{21}|^2 \leq 1$) responses and always $Gain \leq 1$. The gain coefficient has negative values when lasing is achieved and positive values when absorption is dominant. Hence, THz lasing or amplification is always obtained for negative gain values. For the actual calculation of gain values at the divergent points, nonlinear effects stabilizing the lasing should be taken into account, which are out of the scope of the current work and will be considered in the future. Figure 6(a) demonstrates the comparison of the computed gain coefficient obtained by simulations (black dashed line) and the presented in the previous section analytical transmission line method (red solid line) when the quasi-Fermi level is fixed to 50 meV and for the same patterned HMM configuration. The results match very well and the minor resonance frequency shift between the two modeling methods is attributed to the usual limitations of the numerical modeling method due to the used finite size mesh. An excellent agreement exists between the frequency point where the maximum gain is obtained in Fig. 6(a) and the near–zero group velocity frequency range demonstrated before by the dispersion diagram in Fig. 3(b). To further analytically demonstrate and verify the obtained lasing effect, we plot the $|M_{22}|$ coefficient versus the incident frequency and quasi-Fermi level, as shown in Fig. 6(b), where the used parameters are the same with the results presented in Fig. 6(a). We find that $|M_{22}|$ takes close to zero values for a 50 meV quasi-Fermi level, exactly at the gain resonance frequency shown in Fig. 6(a). Interestingly, $|M_{22}|$ has close to zero values across a broad frequency range, demonstrating the tunable response of the proposed device that can lead to THz lasing performance in a broad frequency range. The lasing response demonstrated by the computed $|M_{22}|$ coefficient agrees very well with the transmittance spectrum shown before in Fig. 4(b), as it is expected.

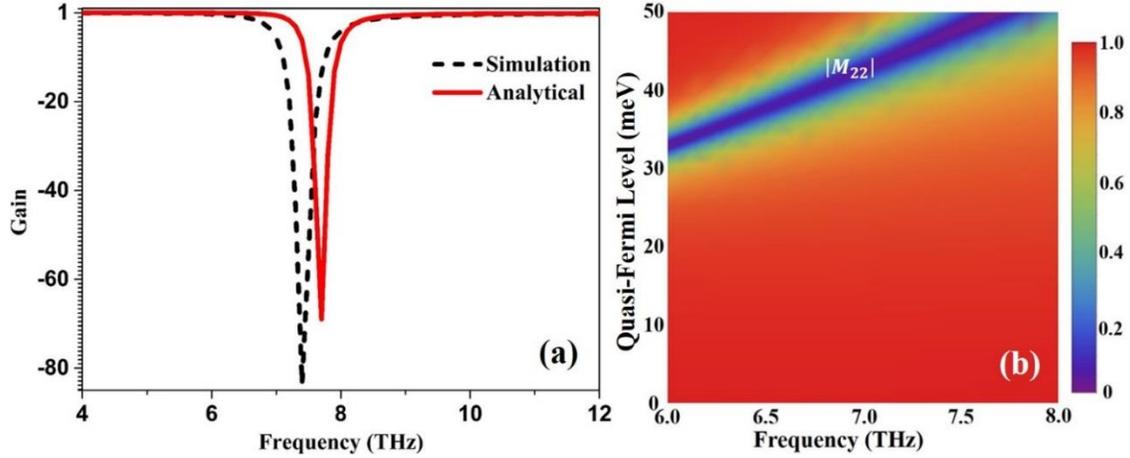

Fig. 6. (a) Simulation and analytical results of the computed gain coefficient in dB of the proposed slow-wave graphene HMM structure. (b) The absolute value of the $M_{22}$ coefficient as a function of the incident frequency and quasi-Fermi level. In both captions, the dimension parameters are fixed to $P = 1.9$ μm, $W = 0.5$ μm, $t = 0.1$ μm. The quasi-Fermi level of the pumped graphene is fixed to 50 meV in caption (a).

Figures 7(a, b) prove that the period and width of the graphene sheets and air slots, respectively, have distinct effects on the amplification of the THz waves. The gain is presented in dB scale and computed by $Gain(dB) = 10\log_{10}(|Gain|)$ (Fig. 7) in order to clearly demonstrate the variation of the gain peak value combined with the lasing frequency shift as a function of the geometrical structure parameters and quasi-Fermi level. We fix the graphene quasi-Fermi level to 50 meV and the air width to 0.5um and a dominant gain peak is reported in Fig. 7(a) for $P = 2.1$ μm. This can also be explained by the lasing relationship between input impedance and free space impedance obtained by the transmission line model, demonstrated before in section 3. Similarly, the gain reaches a peak for the air slot width value $W = 0.5$ μm when the period is fixed to $P = 2.1$ μm and the result can be seen in Fig. 7(b). Strong frequency shift in the resonant THz lasing response of the proposed structure is obtained when its geometrical parameters vary. Finally, we have checked that the proposed structure will also exhibit lasing in a more practical scenario, when it is surrounded by a dielectric substrate and superstrate (see more details in subsequent section 5).

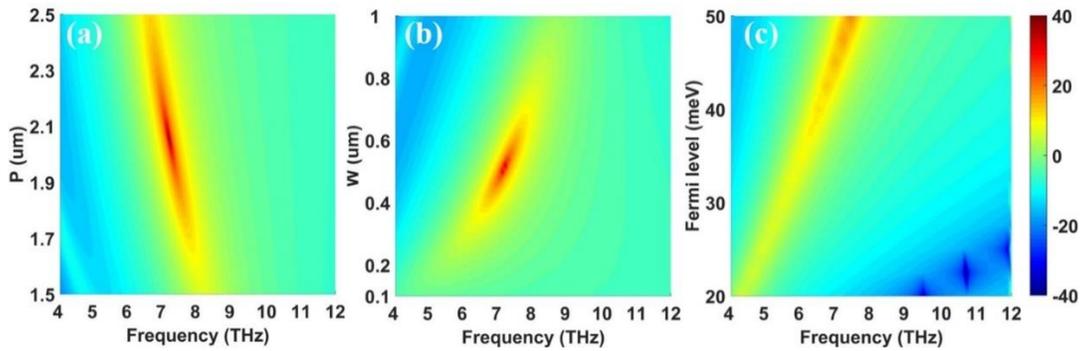

Fig. 7. Contour plots of the gain in dB scale as a function of the (a) period, (b) air width, and (c) quasi-Fermi level versus the frequency of the incoming THz radiation. The graphene quasi-Fermi level is fixed to 50 meV in captions (a) and (b). The air width is fixed to $W = 0.5$ μm in caption (a). The period is fixed to $P = 2.1$ μm in caption (b). The parameters of the slow-wave graphene HMM structure are fixed to $P = 1.9$ μm, $W = 0.5$ μm in caption (c).

Note that tunable THz lasing can interestingly be achieved without changing the geometry of the proposed device and only by modulating the optical pump intensity applied to graphene. The calculated gain coefficient of the currently proposed THz device versus the frequency and quasi-Fermi level is reported in Fig. 7(c). The gain is found to be enhanced as the quasi-Fermi level is increased. As discussed in section 2, the conductivity of the pumped graphene has a strong dependence on its quasi-Fermi level, which determines the electron and hole concentrations. Thus, higher quasi-Fermi level values are translated to stronger population inversion (or lasing) considering that the relaxation time of intraband transitions is always faster than the recombination time of the electron-hole pairs along the pumped graphene monolayers [49]. This leads to more negative graphene conductivity values (larger gain) and is the main reason behind the computed higher gain obtained due to larger quasi-Fermi level values. In addition, the lasing operation frequency blueshifts as the quasi-Fermi level increases, which is also reported in Fig. 7(c), leading to tunable THz lasing. Lastly, it is interesting to note that the proposed lasing effect is temperature-dependent, similar to other THz lasing schemes [40]. The simulation and analytical studies in our work are always performed under a fixed temperature, equal to $T = 77$ K, because higher temperatures will lead to larger losses and lower gain values. The duration and strength of the presented graphene-based THz gain can be extended and improved, respectively, if AB-stacked bilayer graphene [50, 51] is used instead of the current graphene monolayers, which was recently found that it can suppress nonradiative Auger carrier recombination rates due to its large electronic gap [52–54]. Note that the Auger recombination events pose the main limit to the THz gain duration and strength [49], [55], [56]. Finally, we would like to stress that even lasing with femtosecond duration obtained in THz frequencies can be interesting to several applications (for example, sensing).

## 5. Sensing application based on the graphene-HMM structure

Graphene is an ideal material for electrochemical sensing and biosensing since it has large surface-to-volume ratio, excellent electrical conductivity, high carrier mobility, and many other unique properties. As an example, graphene-based highly sensitive chemical and biological sensors have been used to diagnose several diseases [57, 58]. The proposed graphene-HMM structure has great prospect to be used for ultrasensitive sensing applications. The envisioned sensor is composed of a patterned layered graphene-dielectric configuration (10 graphene layers), similar to the currently proposed active slow-wave HMM device, but now embedded in different dielectric materials located at the surrounding space. The schematic of the proposed sensor is presented in the inset of Fig. 8(a). Strong THz amplification and slow-light electromagnetic mode confinement can also be realized with this configuration, similar to the results presented in the previous sections. The gain response of such active structure is expected to be ultrasensitive to the surrounding medium. It is clearly demonstrated in Fig. 8(a) that the loading of different surrounding dielectric materials will lead to a substantial redshift in the THz resonant frequency of the lasing mode. The proposed device is expected to accurately sense various materials with different permittivity values by observing the shift in its resonant frequency. Moreover, we placed a very thin dielectric layer above the active structure to mimic a layer of biomolecules, as shown in the inset of Fig. 8(b). Here, the thickness of the dielectric layer is set to be 100 nm. We find that there is a slight continuous shift to the lasing frequency when we change the relative dielectric constant of this thin layer from 1 to 4, as can be seen in Fig. 8(b). However, we would like to stress that the gap between the patterned graphene-based HMM waveguides is relative large ($W = 0.5$ μm) and it is not sure if the biomolecules will stay on top of the proposed structure or the scenario presented in Fig. 8(a) is more practical. In any case, there is going to be some shift in the resonance frequency of the proposed new sensor design independent of the molecules placement.

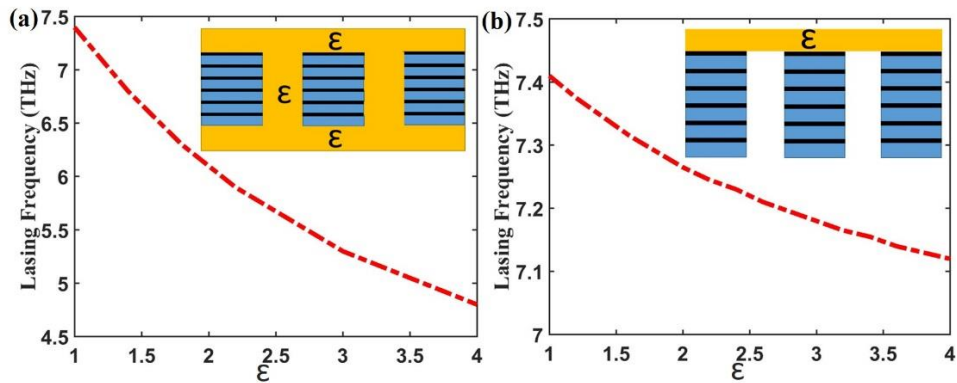

Fig. 8. The lasing frequency as a function of the dielectric constant for the proposed active graphene-based HMM structure (a) embedded in different dielectric materials, and (b) covered with a very thin dielectric layer. Insets: The corresponding schematics of the proposed graphene-based HMM sensor.

## 6. Conclusions

To conclude, a lasing graphene HMM device is proposed based on patterned photopumped graphene-dielectric multilayers. The device is both analytically and numerically investigated. Excellent agreement is found between the results produced by both modeling methods. It is presented that super scattering or coherent lasing emission can occur in the THz frequency range due to the proposed patterned graphene-based HMM device. In addition, the geometrical parameters and quasi-Fermi levels of the presented graphene HMM can strongly affect the THz lasing operation frequency. More interestingly, tunable THz amplification in a broad frequency range is achieved by controlling the optical pumping, without changing the proposed HMM structure dimensions and geometry. The presented device has an extremely subwavelength thickness, making it an ideal candidate to be used as a compact THz source in the envisioned THz on-chip integrated photonic circuits. Finally, we present an interesting THz sensing application based on the proposed active graphene-based HMM device, by demonstrating that the lasing frequency of such a structure is ultrasensitive to the dielectric constant values of different surrounding media.

## Funding

National Science Foundation through the Nebraska Materials Research Science and Engineering Center (MRSEC) (Grant No. DMR-1420645); National Science Foundation (Grant No. ECCS-1711409).

## References

1. A. Poddubny, I. Iorsh, P. Belov, and Y. Kivshar, "Hyperbolic metamaterials," Nat. Photonics **7**(12), 948–957 (2013).
2. K. V. Sreekanth, A. De Luca, and G. Strangi, "Negative refraction in graphene-based hyperbolic metamaterials," Appl. Phys. Lett. **103**(2), 023107 (2013).
3. C. Argyropoulos, N. M. Estakhri, F. Monticone, and A. Alù, "Negative refraction, gain and nonlinear effects in hyperbolic metamaterials," Opt. Express **21**(12), 15037-15407 (2013).
4. A. J. Hoffman, L. Alekseyev, S. S. Howard, K. J. Franz, D. Wasserman, V. A. Podolskiy, E. E. Narimanov, D. L. Sivco, and C. Gmachl, "Negative refraction in semiconductor metamaterials," Nat. Mater. **6**(12), 946–950 (2007).
5. S. Campione, T. S. Luk, S. Liu, and M. B. Sinclair, "Optical properties of transiently-excited semiconductor hyperbolic metamaterials," Opt. Mater. Express **5**(11), 2385 (2015).
6. a V Kabashin, P. Evans, S. Pastkovsky, W. Hendren, G. a Wurtz, R. Atkinson, R. Pollard, V. a Podolskiy, and a V Zayats, "Plasmonic nanorod metamaterials for biosensing," Nat. Mater. **8**(11), 867–871 (2009).
7. C. Guclu, S. Campione, and F. Capolino, "Hyperbolic metamaterial as super absorber for scattered fields generated at its surface," Phys. Rev. B **86**(20), 205130 (2012).
8. J. Zhou, A. F. Kaplan, L. Chen, and L. J. Guo, "Experiment and Theory of the Broadband Absorption by a Tapered Hyperbolic Metamaterial Array," ACS Photonics **1**(7), 618–624 (2014).
9. M. Sakhdari, M. Hajizadegan, M. Farhat, and P.-Y. Chen, "Efficient, broadband and wide-angle hot-electron transduction using metal-semiconductor hyperbolic metamaterials," Nano Energy **26**, 371–381 (2016).
10. P.-Y. Chen, M. Hajizadegan, M. Sakhdari, and A. Alù, "Giant Photoresponsivity of Midinfrared Hyperbolic Metamaterials in the Photon-Assisted-Tunneling Regime," Phys. Rev. Appl. **5**(4), 041001 (2016).
11. D. Lu, J. J. Kan, E. E. Fullerton, and Z. Liu, "Enhancing spontaneous emission rates of molecules using nanopatterned multilayer hyperbolic metamaterials," Nat. Nanotechnol. **9**(1), 48–53 (2014).
12. M. A. Noginov, H. Li, Y. A. Barnakov, D. Dryden, G. Nataraj, G. Zhu, C. E. Bonner, M. Mayy, Z. Jacob, and E. E. Narimanov, "Controlling spontaneous emission with metamaterials," Opt. Lett. **35**(11), 1863-1865 (2010).
13. H. N. S. Krishnamoorthy, Z. Jacob, E. Narimanov, I. Kretzschmar, and V. M. Menon, "Topological Transitions in Metamaterials," Science **336**(6078), 205–209 (2012).
14. T. Li and J. B. Khurgin, "Hyperbolic metamaterials: beyond the effective medium theory," Optica **3**(12), 1388 (2016).
15. Y. Guo, W. Newman, C. L. Cortes, and Z. Jacob, "Applications of Hyperbolic Metamaterial Substrates," Adv. Optoelectron. **2012**(1), 1–9 (2012).
16. A. K. Geim and K. S. Novoselov, "The rise of graphene," Nat. Mater. **6**(3), 183–191 (2007).
17. P.-Y. Chen, J. Soric, Y. R. Padooru, H. M. Bernety, A. B. Yakovlev, and A. Alù, "Nanostructured graphene metasurface for tunable terahertz cloaking," New J. Phys. **15**(12), 123029 (2013).
18. P.-Y. Chen and A. Alu, "Terahertz Metamaterial Devices Based on Graphene Nanostructures," IEEE Trans. Terahertz Sci. Technol. **3**(6), 748–756 (2013).
19. B. Jin, T. Guo, and C. Argyropoulos, "Enhanced third harmonic generation with graphene metasurfaces," J. Opt. **19**(9), 094005 (2017).
20. M. A. K. Othman, C. Guclu, and F. Capolino, "Graphene-based tunable hyperbolic metamaterials and enhanced near-field absorption," Opt. Express **21**(6), 7614-7632 (2013).
21. Y.-C. Chang, C.-H. Liu, C.-H. Liu, S. Zhang, S. R. Marder, E. E. Narimanov, Z. Zhong, and T. B. Norris, "Realization of mid-infrared graphene hyperbolic metamaterials," Nat. Commun. **7**, 10568 (2016).
22. I. V. Iorsh, I. S. Mukhin, I. V. Shadrivov, P. A. Belov, and Y. S. Kivshar, "Hyperbolic metamaterials based on multilayer graphene structures," Phys. Rev. B **87**(7), 075416 (2013).
23. M. A. K. Othman, C. Guclu, and F. Capolino, "Graphene–dielectric composite metamaterials: evolution from elliptic to hyperbolic wavevector dispersion and the transverse epsilon-near-zero condition," J. Nanophotonics **7**(1), 073089 (2013).
24. L. Ju, B. Geng, J. Horng, C. Girit, M. Martin, Z. Hao, H. a Bechtel, X. Liang, A. Zettl, Y. R. Shen, and F. Wang, "Graphene plasmonics

25. A. N. Grigorenko, M. Polini, and K. S. Novoselov, "Graphene plasmonics," Nat. Photonics **6**(11), 749–758 (2012).
26. T. Guo and C. Argyropoulos, "Broadband polarizers based on graphene metasurfaces," Opt. Lett. **41**(23), 5592-5595 (2016).
27. P.-Y. Chen, C. Argyropoulos, M. Farhat, and J. S. Gomez-Diaz, "Flatland plasmonics and nanophotonics based on graphene and beyond," Nanophotonics **6**(6), 1239–1262 (2017).
28. C. Argyropoulos, "Enhanced transmission modulation based on dielectric metasurfaces loaded with graphene," Opt. Express **23**(18), 23787 (2015).
29. T. Gric and O. Hess, "Tunable surface waves at the interface separating different graphene-dielectric composite hyperbolic metamaterials," Opt. Express **25**(10), 11466-11476 (2017).
30. J. M. Hamm, A. F. Page, J. Bravo-Abad, F. J. Garcia-Vidal, and O. Hess, "Nonequilibrium plasmon emission drives ultrafast carrier relaxation dynamics in photoexcited graphene," Phys. Rev. B **93**(4), 041408 (2016).
31. P.-Y. Y. Chen and J. Jung, "PT Symmetry and Singularity-Enhanced Sensing Based on Photoexcited Graphene Metasurfaces," Phys. Rev. Appl. **5**(6), 064018 (2016).
32. P. Weis, J. L. Garcia-Pomar, and M. Rahm, "Towards loss compensated and lasing terahertz metamaterials based on optically pumped graphene," Opt. Express **22**(7), 8473-8489 (2014).
33. V. Ryzhii, M. Ryzhii, and T. Otsuji, "Negative dynamic conductivity of graphene with optical pumping," J. Appl. Phys. **101**(8), 083114 (2007).
34. V. V. Popov, O. V. Polischuk, A. R. Davoyan, V. Ryzhii, T. Otsuji, and M. S. Shur, "Plasmonic terahertz lasing in an array of graphene nanocavities," Phys. Rev. B - Condens. Matter Mater. Phys. **86**(19), 195437 (2012).
35. V. Ryzhii, M. Ryzhii, V. Mitin, and T. Otsuji, "Toward the creation of terahertz graphene injection laser," J. Appl. Phys. **110**(9), 094503 (2011).
36. T. Watanabe, T. Fukushima, Y. Yabe, S. A. Boubanga Tombet, A. Satou, A. A. Dubinov, V. Y. Aleshkin, V. Mitin, V. Ryzhii, and T. Otsuji, "The gain enhancement effect of surface plasmon polaritons on terahertz stimulated emission in optically pumped monolayer graphene," New J. Phys. **15**(7), 075003 (2013).
37. T. Otsuji, T. Watanabe, S. A. Boubanga Tombet, A. Satou, W. M. Knap, V. V Popov, M. Ryzhii, and V. Ryzhii, "Emission and Detection of Terahertz Radiation Using Two-Dimensional Electrons in III–V Semiconductors and Graphene," IEEE Trans. Terahertz Sci. Technol. **3**(1), 63–71 (2013).
38. D. Svintsov, V. Ryzhii, A. Satou, T. Otsuji, and V. Vyurkov, "Carrier-carrier scattering and negative dynamic conductivity in pumped graphene," Opt. Express **22**(17), 19873-19886 (2014).
39. M. Tonouchi, "Cutting-edge terahertz technology," Nat. Photonics **1**(2), 97–105 (2007).
40. M. S. Vitiello, G. Scalari, B. Williams, and P. De Natale, "Quantum cascade lasers: 20 years of challenges," Opt. Express **23**(4), 5167-5182 (2015).
41. O. Kidwai, S. V. Zhukovsky, and J. E. Sipe, "Effective-medium approach to planar multilayer hyperbolic metamaterials: Strengths and limitations," Phys. Rev. A **85**(5), 053842 (2012).
42. O. Kidwai, S. V. Zhukovsky, and J. E. Sipe, "Dipole radiation near hyperbolic metamaterials: applicability of effective-medium approximation," Opt. Lett. **36**(13), 2530-2532 (2011).
43. L. A. Falkovsky and A. A. Varlamov, "Space-time dispersion of graphene conductivity," Eur. Phys. J. B **56**(4), 281–284 (2007).
44. A. Fallahi and J. Perruisseau-Carrier, "Design of tunable biperiodic graphene metasurfaces," Phys. Rev. B **86**(19), 195408 (2012).
45. N. Marcuvitz, Waveguide Handbook (McGraw-Hill, New York, 1951).
46. D. Pozar, *Microwave Engineering Fourth Edition* (2005).
47. Y. D. Chong, L. Ge, H. Cao, and A. D. Stone, "Coherent Perfect Absorbers: Time-Reversed Lasers," Phys. Rev. Lett. **105**(5), 053901 (2010).
48. Comsol Multiphysics, http://www.comsol.com/.
49. I. Gierz, J. C. Petersen, M. Mitrano, C. Cacho, I. C. E. Turcu, E. Springate, A. Stöhr, A. Köhler, U. Starke, and A. Cavalleri, "Snapshots of non-equilibrium Dirac carrier distributions in graphene," Nat. Mater. **12**(12), 1119–1124 (2013).
50. I. Gierz, M. Mitrano, J. C. Petersen, C. Cacho, I. C. Edmond Turcu, E. Springate, A. Stöhr, A. Köhler, U. Starke, and A. Cavalleri, "Population inversion in monolayer and bilayer graphene," J. Phys. Condens. Matter **27**(16), 164204 (2015).
51. T. Low, P.-Y. Chen, and D. N. Basov, "Superluminal plasmons with resonant gain in population inverted bilayer graphene," Phys. Rev. B **98**(4), 041403 (2018).
52. Y. Zhang, T.-T. Tang, C. Girit, Z. Hao, M. C. Martin, A. Zettl, M. F. Crommie, Y. R. Shen, and F. Wang, "Direct observation of a widely tunable bandgap in bilayer graphene," Nature **459**(7248), 820–823 (2009).
53. D. M. Basko, "Theory of resonant multiphonon Raman scattering in graphene," Phys. Rev. B **78**(12), 125418 (2008).
54. K. T. Delaney, P. Rinke, and C. G. Van de Walle, "Auger recombination rates in nitrides from first principles," Appl. Phys. Lett. **94**(19), 191109 (2009).
55. E. Malic, T. Winzer, F. Wendler, and A. Knorr, "Review on carrier multiplication in graphene," Phys. status solidi **253**(12), 2303–2310 (2016).
56. G. Alymov, V. Vyurkov, V. Ryzhii, A. Satou, and D. Svintsov, "Auger recombination in Dirac materials: A tangle of many-body effects," Phys. Rev. B **97**(20), 205411 (2018).
57. B. Zhang and T. Cui, "An ultrasensitive and low-cost graphene sensor based on layer-by-layer nano self-assembly," Appl. Phys. Lett. **98**(7), 073116 (2011).
58. H. Jiang, "Chemical Preparation of Graphene-Based Nanomaterials and Their Applications in Chemical and Biological Sensors," Small **7**(17), 2413-2427 (2011).